**Tunable Ferromagnetism in LaCoO$_3$ Epitaxial Thin Films**

*Dongwon Shin[#], Sangmoon Yoon[#], Sehwan Song, Sungkyun Park, Ho Nyung Lee, and Woo Seok Choi[*]*

D. Shin, W. S. Choi

Department of Physics, Sungkyunkwan University, Suwon 16419, Republic of Korea

E-mail: choiws@skku.edu

S. Yoon, H. N. Lee

Materials Science and Technology Division, Oak Ridge National Laboratory, Oak Ridge, Tennessee 37831, United States

S. Yoon

Department of Physics, Gachon University, Seongnam, 13120, Republic of Korea

S. Song, S. Park

Department of Physics, Busan National University, Busan 46241, Republic of Korea

Ferromagnetic insulators play a crucial role in the development of low-dissipation quantum magnetic devices for spintronics. Epitaxial LaCoO$_3$ thin film is a prominent ferromagnetic insulator, in which the robust ferromagnetic ordering emerges owing to epitaxial strain. Whereas it is evident that strong spin-lattice coupling induces ferromagnetism, the reported ferromagnetic properties of epitaxially strained LaCoO$_3$ thin films were highly consistent. For example, even under largely modulated degree of strain, the reported Curie temperatures of epitaxially strained LaCoO$_3$ thin films lie within 80–85 K, without much deviation. In this study, substantial enhancement (~18%) in the Curie temperature of epitaxial LaCoO$_3$ thin films is demonstrated via crystallographic orientation dependence. By changing the crystallographic orientation of the films from (111) to (110), the crystal-field energy was reduced and the charge transfer between the Co and O orbitals was enhanced. These modifications led to a considerable enhancement of the ferromagnetic properties (including the Curie temperature and magnetization), despite the identical nominal degree of epitaxial strain. The findings of this study provide insights into facile tunability of ferromagnetic properties via structural symmetry control in $LaCoO_3$.

## 1. Introduction

Ferromagnetic insulators (FM-Is) are essential components in dissipation-less spintronic devices with highly suppressed leakage currents.[1, 2] They generate pure spin currents by filtering charges and introducing spins into adjacent nonmagnetic layers via the magnetic proximity effect.[3-8] However, FM-Is are scarce in nature compared to more conventional FM metals because the general superexchange mechanism in insulators promotes antiferromagnetic (AFM) ordering.[9-12] Iron yttrium garnet (YIG), $\alpha$-$Fe_2O_3$, spinel $CoFe_2O_4$, EuS, EuO, and strained $EuTiO_3$ thin films are some of examples of FM-I materials.[13-19] Epitaxially-strained $LaCoO_3$ (LCO), e.g., epitaxially deposited on $SrTiO_3$ (STO), $(LaAlO_3)_{0.3}(Sr_2AlTaO_8)_{0.7}$ (LSAT), or $YAlO_3$ substrates, is an emerging FM-I with a Curie temperature above the liquid $N_2$ temperature ($T_C$ = ~80 K), with a robust ferromagnetic origin.[20, 21]

**Table 1. $T_C$ obtained in LCO thin films:** A summary of reported $T_C$ of LCO thin films, depending on the applied magnetic field ($H$), type of substrates, crystallographic orientation, and film thickness. Note that a robust $T_C$ with few exceptions is found for the LCO thin films. In contrast a substantial increase of $T_C$ (~ 18%) is observed in the current study.

| $T_C$ (K) | $H$ [Oe] | substrate | orientation | thickness [nm] | reference |
|---|---|---|---|---|---|
| 43.2 | 200 | LAO | (100) | 100 | |
| 69.2 | 200 | SLAO | (100) | 100 | |
| 71.6 | 200 | SLGO | (100) | 100 | *Phys. Rev. B* **77**, 014434 (2008) |
| 76.0 | 200 | LSAT | (100) | 100 | |
| 77.6 | 200 | STO | (100) | 100 | |
| 80/87 | 200 | LSAT | (100) | 8/50 | *Phys. Rev. B* **79**, 024424 (2009) |
| 86 | 2000 | STO | (100) | 100 | |
| 75 | 2000 | LAO | (100) | 100 | |
| 85 | 2000 | LSAT | (100) | 100 | *Eur. Phys. J. B* **76**, 215–219 (2010) |
| 87 | 2000 | PMN-PT | (100) | 100 | |
| 84 | 2000 | SLAO | (100) | 100 | |
| 80 | 2000 | STO/LSAT | (100) | 20-90 | *J. Appl. Phys.* **105**, 07E503 (2009) |
| 76-83 | 50 | LAO | (100) | 30-95 | *J. Appl. Phys.* **109**, 07D717 (2011) |
| 75.9/81.2 | 50 | STO/LSAT | (100) | 15 | *Phys. Rev. B* **91**, 144418 (2015) |
| 80 | 1000 | STO/LSAT | (100) | 30 | *Nano Lett.* **12**, 4966-4970 (2012) |
| 94 | 1000 | LSAT | (110) | 60 | *Phys. Rev. Lett.* **111**, 027206 (2013) |
| 24 | 1000 | LSAT | (111) | 60 | *Phys. Rev. B* **92**, 195115 (2015) |
| 92/90 | 500 | STO/LSAT | (111)/(100) | 35/38 | |
| 85 | 500 | STO/LSAT | (100) | 30 | *AIP Advances* **8**, 056317 (2018) |
| 85 | 500 | STO | (100) | 12 | *PNAS* **115**, 2873–2877 (2018) |
| 78 | 500 | STO | (100) | 70 | *Curr. Appl. Phys.* **28**, 87-92 (2021) |
| 50 | 500 | STO | (111) | 80 | *Phys. Status Solidi B*, 2100424 (2021) |
| 76.1-90.2 | 1000 | STO | (100), (110) and (111) | 30 | This work |

Long-range FM ordering in epitaxial LCO thin films is facilitated by strain-induced lattice distortions. Although bulk LCO is nonmagnetic with a rhombohedral crystal structure, the tensile strain in the thin film is known to induce monoclinic and tetragonal phases, resulting in a spin-ordered state.[20, 22, 23] The large and facile modification of the crystalline symmetry originates from the ferroelastic nature of LCO, which favors the formation of twin-domain structures under mechanical pressure or epitaxial strain.[24-31] Such twin-domain structures have been observed as dark stripe patterns in high-angle annular dark-field scanning transmission electron microscopy (HAADF STEM) imaging.[20, 23, 32, 33] Detailed microstructure analyses using HAADF STEM on (100)-oriented LCO thin films enable classification of basic lattice units into compressed- (*c*-unit), tensile-stretched- (*t*-unit), and bulk-units (*b*-unit).[23] Among these, *c*-units with coexisting low (LS) and high spin- (HS) states of the Co spin configuration were found to be responsible for the FM phase. It should be noted that the FM phase, characterized by $T_C$, in the (100)-oriented LCO thin film was mostly unchanged. As summarized in **Table 1**, the reported values of $T_C$ of (100)-oriented LCO thin films are mostly in the range 80–85 K,[20, 34-45] despite the large difference in the degree of epitaxial strain. Note that $T_C$ might decrease further below 80 K, if the crystalline quality is not ideal.[43] Such unsusceptible behavior in $T_C$ suggests that the magnetically active *c*-units are highly resilient against orthogonal stress, likely stemming from the ferroelastic nature of LCO.[23]

Modifications in the crystallographic orientation lead to disparate lattice symmetry, offering extra tunability in the functional properties of epitaxial thin films. Even though an identical film/substrate system is used, implying the same degree of nominal epitaxial strain, the structural symmetry of the thin film can be modulated by the crystallographic orientation of the substrate, leading to stress that is not orthogonal to the lattice. For instance, the FM $T_C$ of epitaxial $SrRuO_3$ thin films can be increased from 147 to 157 K, as the crystallographic orientation changes from (100) to (111), because of the decreased spin dimensionality within the trigonal symmetry.[46] In epitaxial $La_{0.7}Sr_{0.3}MnO_3$ thin films, (110) and (111)-oriented thin films exhibit larger spin moments than (100)-oriented thin films.[47] As the crystallographic orientation of $LaAlO_3$ thin films on $KTaO_3$ substrates changed from (100) to (110) or (111), emergent superconductivity was observed. The superconducting critical temperatures could be further modulated between the (110) and (111)-oriented thin films, with the values of 1 and 2 K, respectively.[48-50] Because the crystallographic orientation determines the crystallographic symmetry of the lattice unit, it is expected to modify the robust FM-I phase of LCO epitaxial thin films as well. This strategy would provide fundamental insight into the structure-property relationship of LCO, which would not have been possible using only (100)-oriented thin films with different strain states, but the same type of orthogonal stress. Indeed, studies on LCO thin films with crystallographic orientations other than (100)-orientation are highly limited and report inconsistent $T_C$ values, as shown in Table 1, and a systematic comparison between (100)-, (110)-, and (111)-oriented LCO epitaxial thin films has not been conducted thus far.

In this study, we demonstrate a large modulation of the FM phase in LCO epitaxial thin films, including the substantial tuning of $T_C$, by tailoring the crystalline symmetry via crystallographic orientation control. We fabricated epitaxial LCO thin films on (100)-, (110)- and (111)-oriented STO substrates using pulsed laser epitaxy (PLE). The structural quality and crystallographic symmetry of the thin films have been determined by X-ray diffraction (XRD) measurements and HAADF STEM imaging. The FM phases for the (100)-, (110)- and (111)-oriented LCO thin films have $T_C$s of 81.5, 90.2, 76.1 K, respectively. The large modulation of the FM $T_C$, i.e., ~18% enhancement in the (110)-oriented thin film compared to the (111)-oriented one, captures the effectiveness of the crystallographic symmetry control on the lattice units. The physical origin of the $T_C$ enhancement was investigated by characterizing the microscopic lattice units, electronic structure measurements, and theoretical calculations. In particular, the reduced crystal-field splitting energy and stronger Co-O charge transfer in the (110)-oriented thin film resulted in considerable enhancement of the FM properties.

## 2. Results and Discussion

The global lattice structures of the LCO thin films (30 nm) were characterized using XRD to reveal the systematic crystallographic orientation dependence. **Figure 1**a shows the XRD $\theta-2\theta$ scans for the LCO thin films. Fine thickness fringes are commonly observed along with the corresponding LCO peaks (Figure 1a; Figure S1a and b, Supporting Information), indicating consistent crystalline phases with atomically smooth surfaces and interfaces. In addition, atomic force microscopy images show atomically smooth surfaces of the epitaxial LCO thin films with low surface roughness of < 1 u.c. (Figure S2, Supporting Information).

Figure 1b shows the XRD reciprocal space maps around the (103), (120), and (212) Bragg reflections for the LCO thin films epitaxially grown on the (100)-, (110)-, and (111)-oriented STO substrates, respectively. Despite the difference in the crystallographic orientations, all films were fully strained to the substrates. Although the nominal lattice mismatch between the thin film and the substrate is identical, it is important to note that the degree of strain still differs owing to symmetry modification. The difference in the degree of strain is shown in Figures 1c, in terms of the pseudocubic unit cell volume ($V_{u.c.}$) and out-of-plane stress ($\varepsilon_{oop}$ (%) = 100 × ($d_{film}$ − $d_{bulk}$) / $d_{bulk}$, where $d_{film}$ and $d_{bulk}$ are the out-of-plane distances between the atomic planes of the thin film and the bulk, respectively) for distinctive crystallographic orientations. The same trends are shown for the 10- and 20-nm-thick LCO thin films (Figure S1c and d, Supporting Information). Both $V_{u.c.}$ and $\varepsilon_{oop}$ were the largest for the (110)-oriented thin film, indicating the largest lattice distortion, and the smallest for the (111)-oriented film.

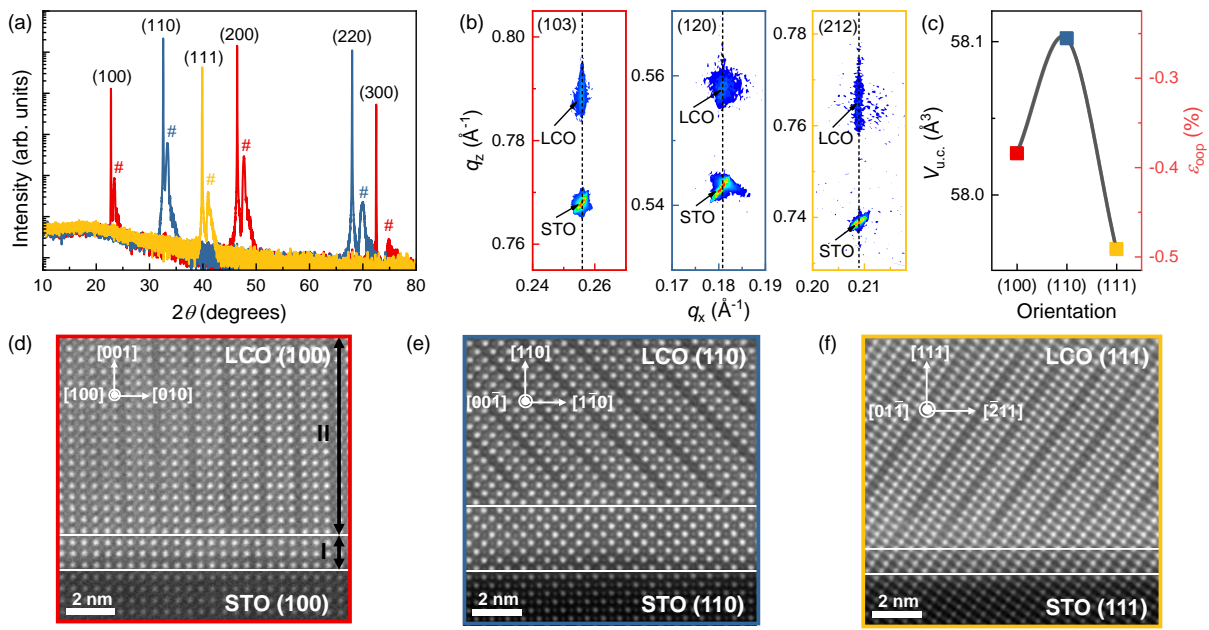

**Figure 1. Crystal structures of epitaxial LCO thin films with various crystallographic orientations:** a) XRD $\theta$-$2\theta$ scans of epitaxial LCO thin films (#) grown on STO substrates with different orientations. b) XRD reciprocal space maps (RSM) of the (100)-, (110)-, and (111)-oriented LCO thin films, shown around the (103), (120), and (212) Bragg reflections of the STO substrates, respectively. Crystallographic orientation-dependent c) unit cell volume ($V_{u.c.}$) and out-of-plane strain ($\varepsilon_{oop}$). Z-contrast HAADF STEM images for LCO thin films grown on d) (100), e) (110), and f) (111) STO substrates. Dark stripe patterns are observed in all crystallographic orientations of the LCO thin films. Two distinct regions are marked with white horizontal lines. Region I near the interface is a "uniform region" without any dark

stripe patterns. Region II away from the interface is an "ordered region" with dark stripe patterns.

Figure 1d–f shows high-resolution HAADF STEM images of epitaxial LCO thin films grown on d) (100)-, e) (110)-, and f) (111)-oriented STO substrates, respectively. The STEM images commonly exhibit a coherent alignment of the atomic columns without any defects and/or misfit dislocations at the interfaces. Interestingly, distinctive dark stripe patterns (with ~3 u.c. periodicity) were observed, above a certain thickness of the interfacial layer. We carefully exclude the presence of oxygen vacancy ordering at the dark stripe patterns by electron energy loss spectroscopy (EELS) measurements (Figure S3, Supporting Information). While the dark stripe patterns have been frequently recognized for the (100)-oriented LCO thin films,[20, 23, 32, 33] we report similar patterns in (110)- and (111)-oriented thin films with systematic changes for the first time. The LCO thin films can be categorized into two regions considering the dark stripe pattern, i.e., Region I (near the interface) and Region II (above the interface).[20] Region I is free of the dark stripe patterns, suggesting a conventional fully strained state. The thickness of Region I is the thickest for the (110)-oriented thin film, and the thinnest for the (111)-oriented thin film, consistent with the fact that the (110)-oriented thin film has the largest deviation in $V_{u.c.}$ from the bulk value (55.604 Å$^3$). On the other hand, Region II presents twin-domain structures due to the ferroelastic nature of the LCO thin film, as mentioned earlier. For more detailed structural analyses, we measured the distance of La columns in Region II along the two orthogonal axes perpendicular to the HAADF STEM image (Table S1, Supporting Information). The lattice structure of magnetically active $c$-units, which was robust in the (100)-oriented LCO thin films,[21] is distorted differently depending on the substrate orientations, suggesting that the robust structure can be tuned by symmetry constraints. The volume of $c$-unit is the largest in the (110)-oriented LCO thin film, while it is the smallest on the (111)-oriented film, which is consistent with the XRD results.

The FM ordering in the LCO thin films is strongly coupled to the crystallographic symmetry. **Figure 2**a shows the temperature-dependent magnetization, $M(T)$, of the LCO thin films under field-cooled (FC, 1000 Oe) cooling along the in-plane direction. A clear FM transition with a sudden upturn of $M(T)$ is observed at 81.5, 90.2, and 76.1 K for the (100)-, (110)-, and (111)-oriented LCO thin films, respectively. The inset of Figure 2a shows the linear extrapolation of $M(T)$ for the $T_C$ determination. The magnetic-field-dependent magnetization, $M(H)$, measured at 2 K, demonstrates a clear FM hysteresis loop, as shown in Figure 2b. The values of saturation magnetization ($M_s$) and remanent magnetization ($M_r$) are summarized in Figure 2c and 2d, respectively, exhibiting consistent orientation-dependent trends with $T_C$ (Figure 2e). Indeed, a substantial enhancement in ferromagnetic properties, i.e., $T_C$ (~18%), $M_s$ (~36%), and $M_r$ (~49%), is observed, by comparing the (110)-oriented thin film to the (111)-oriented one. The same trend was observed from the Curie-Weiss (C-W) law fitting of the $M(T)$ curves, in terms of the effective ratio of the HS-state (Figure S4, Supporting Information).[29] Within the prevailing spin configuration model of LCO epitaxial thin films,[23, 29, 33, 51] in which the LS- and HS-states coexist, the effective ratios of HS-state are 68.4, 69.5, and 57.2%, respectively, for the (100)-, (110)-, and (111)-oriented LCO thin films. X-ray photoelectron spectroscopy results consistently support that the (110)-oriented LCO thin film exhibits the largest HS state ratio (Figure S5, Supporting Information).

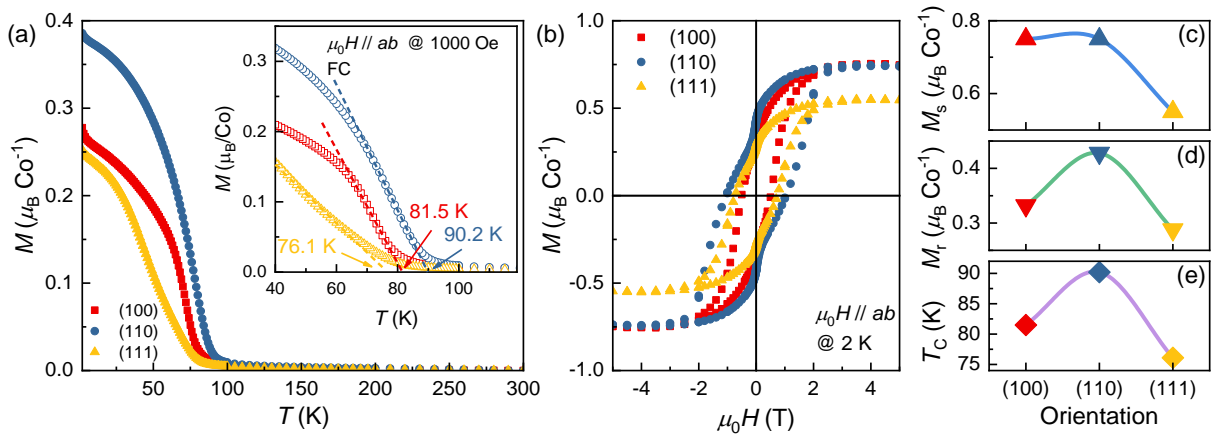

**Figure 2. Crystallographic orientation dependent magnetic properties:** a) $M(T)$ curves of LCO thin films on (100), (110), and (111) STO substrates along the in-plane direction. $M(T)$ curves were obtained during field-cooled cooling at 1000 Oe of magnetic field. The inset of Figure 2a shows $T_C$'s defined by linear extrapolation. b) $M(H)$ curves of LCO thin films on (100), (110), and (111) STO substrates along the in-plane direction at 2 K. $M(H)$ curves of LCO thin films show FM hysteresis loops. Crystallographic orientation-dependent FM properties including c) saturation magnetization ($M_s$), d) remanent magnetization ($M_r$), and e) Curie temperature ($T_C$) are summarized for the LCO thin films.

The electronic structures of LCO thin films provide a microscopic mechanism for FM modulation with distinctive crystallographic symmetries. In particular, we performed spectroscopic ellipsometry to capture minute changes in the electronic band structures of the LCO thin films. The real part of the optical conductivity, $\sigma_1(\omega)$, of the (100)-, (110)-, and (111)-oriented LCO thin films is shown in **Figure** 3a, which closely resembles $\sigma_1(\omega)$ of bulk

LCO.[20, 52] Below 4 eV, $\sigma_1(\omega)$ consists of two Lorentz oscillators centered at ~1.5, and ~3.0 eV, labelled as the $\beta$ (*d-d* transition within Co 3*d* orbitals, brown dotted lines) and $\gamma$ peaks (charge transfer transition, O 2*p* → Co 3*d*, green dotted lines), respectively, as shown in Figure 3b.[52] The quantitative values of the peak positions ($\omega_j$) and spectral weights ($SW_j$) of the $\beta$ and $\gamma$ peaks are summarized in Figure 3c and 3d. The (110)-oriented LCO thin film exhibited a lower $\omega_j$ for both the $\beta$ and $\gamma$ transitions compared to those of the (100)- and (111)-oriented thin films. In addition, the (110)-oriented thin film showed a larger $SW_j$ value for both the $\beta$ and $\gamma$ transitions compared to those of the (100)- and (111)-oriented thin films.

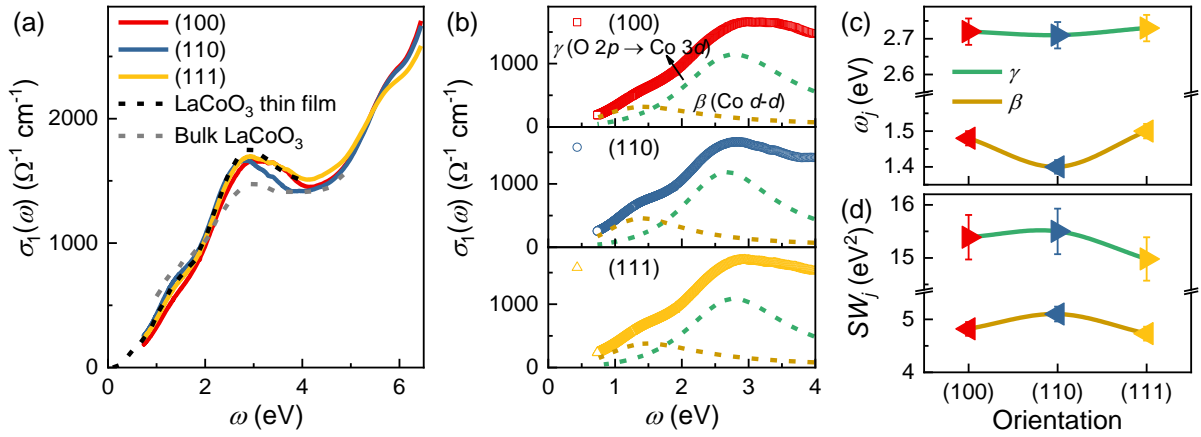

**Figure 3. Optical properties and electronic structures:** a) $\sigma_1(\omega)$ of LCO thin films on differently oriented STO substrates. For comparison, we displayed the reported $\sigma_1(\omega)$ of LCO thin film (black dotted line) and bulk (grey dotted line). b) $\sigma_1(\omega)$ can be deconvoluted by the Lorentzian oscillators corresponding to $\beta$ (Co *d-d* transition) and $\gamma$ (O 2*p* to Co 3*d* charge transfer transition) peaks. Each c) $\omega_j$ and d) $SW_j$ of the individual Lorentzian oscillators is shown as a function of the crystallographic orientation of the LCO thin films.
LCO thin film: Reproduced with permission.[20] Copyright 2012, ACS Publications. Bulk LCO: Reproduced with permission.[52] Copyright 2014, Nature portfolio.

The strongly intertwined electronic structure and FM properties of the LCO epitaxial thin films modulated by crystallographic orientation are illustrated in **Figure 4.** When we consider an HS configuration within LCO, which is necessary for FM ordering, the net spin moment can be evaluated from the energy states of the electron orbitals within the atomic picture. In particular, this tendency can be qualitatively captured by taking into account the energy cost ($\Delta$) of the spin-state transition from an LS to HS configuration. A larger spin moment, and hence stronger FM behavior, is expected for a smaller $\Delta$. $\Delta$ can be defined as, $\Delta = \Delta_{CF} - \Delta_{ex} - W/2$, where $\Delta_{CF}$ and $\Delta_{ex}$ are the crystal-field splitting and Hund energies, respectively, and $W$ is the bandwidth.[29] Thus, the competition between $\Delta_{CF}$, $\Delta_{ex}$, and $W$ determines $\Delta$. Both the bond length ($r_{Co-O}$, average distance between Co and O ions) and bond angle ($\theta$, Co-O-Co angle) influence $\Delta$ and the energy scales,[51, 53, 54] in which $\Delta_{CF} \sim r_{Co-O}^{-5}$ and $W \sim r_{Co-O}^{-3.5}\cos(\pi - \theta)$.[29, 55] Apparently, $\Delta_{CF}$ has a stronger dependence on $r_{Co-O}$, and hence, is a dominant factor for determining $\Delta$. For the CoO$_6$ octahedra, $\Delta_{CF}$ is determined by the energy difference between the Co $t_{2g}$ and $e_g$ states, as depicted in Figure 4a. Because $\Delta_{CF}$ is directly related to $\omega_\beta$ and partially related to $\omega_\gamma$, the observation of the lowest $\omega_\beta$ and $\omega_\gamma$ in the (110)-oriented LCO thin film (Figure 3c) indicates that the film has the lowest $\Delta_{CF}$, leading to a stronger FM spin ordering (small $\Delta$, Figure 2).[55, 56] We can directly infer this insight from the lattice structure, as the (110)-oriented LCO thin film has the largest $V_{u.c.}$ among the films studied (Figures 1c and 1d; Table S1, Supporting Information). We further note that the charge transfer transition from the O 2$p$ to Co 3$d$ orbitals, depicted by $SW_\gamma$, is the largest for the (110)-oriented LCO thin film. We expect that (110)-oriented LCO thin film to have $\theta$ value closer to 180° than the other orientations, as it has the largest $V_{u.c.}$, that would lead to a stronger charge transfer between Co and O orbitals.[34, 35] It has been reported that a stronger charge transfer transition leads to enhanced magnetic ordering in LCO thin films.[34, 57] In general, the facile spin and charge transfer between Co 3$d$ and O 2$p$ orbitals introduces a strong FM superexchange interaction.[34, 58]

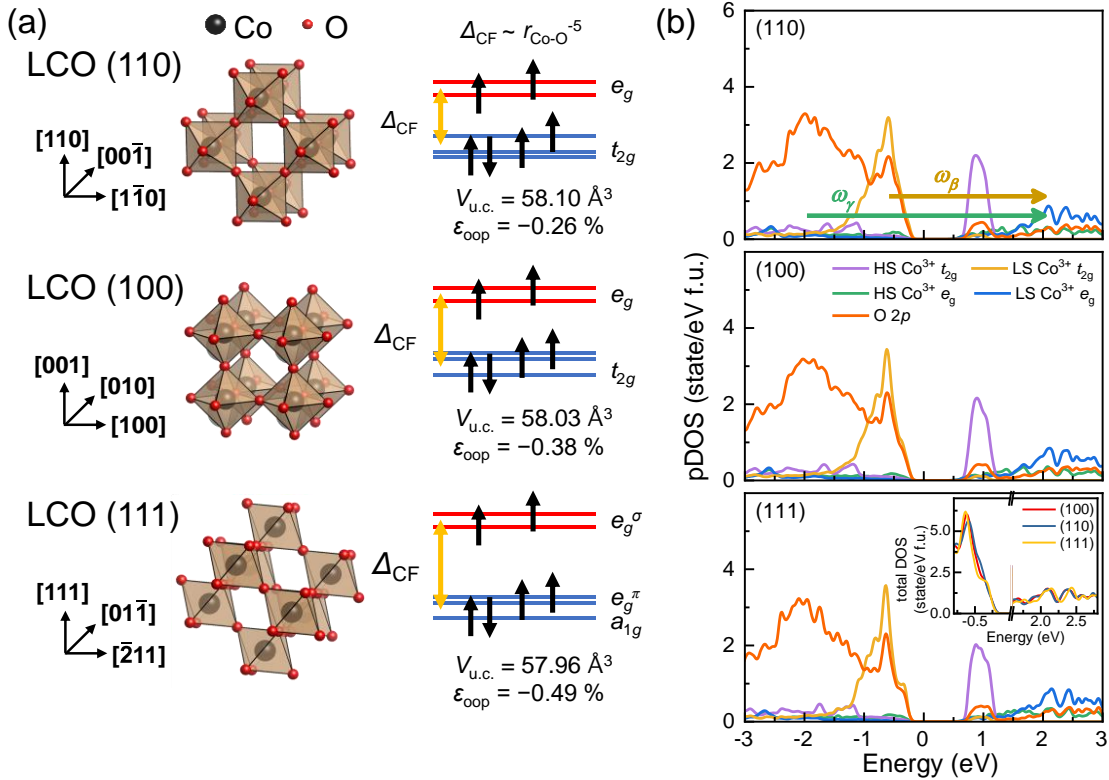

**Figure 4. Origin of the FM modulation in the LCO thin films:** a) Schematic diagram of the crystallographic orientation-dependent spin-state configuration. Crystallographic symmetry modulates the orbital energy level for the $Co^{3+}$ state to determine $\Delta_{CF}$. b) The pDOS and total DOS (inset of bottom panel) of LCO thin films under different symmetry obtained by DFT calculation. Optical transitions $\beta$ (brown arrow) and $\gamma$ (green arrow) indicate the Co *d-d* transition and Co-O charge transfer transition, respectively.

The total (total DOS) and partial density of states (pDOS) calculated by *ab initio* density functional theory (DFT) calculations consistently support the modulation in $\Delta_{CF}$ and charge transfer between the Co and O orbitals, as shown in Figure 4b. We used the experimental lattice parameters obtained from the HAADF STEM analyses (Table S1, Supporting Information) for the calculations. Previously, it was found that the rock salt-type HS/LS-FM order within the *c*-unit is responsible for the magnetic properties of epitaxially strained LCO.[23, 54] Therefore, we used the rock salt-type HS/LS-FM order, which is the most stable state regardless the crystallographic symmetry (Table S2, Supporting Information). The modulated optical transitions $\beta$ (brown arrow) and $\gamma$ (green arrow) from the DFT calculations are systematically reproduced, as shown in the inset of the bottommost panel of Figure 4b. As the crystallographic orientation changes from (111) to (100) to (110), the separation between the Co-O hybridized states and the Co 3*d* states increases on the order of 0.1 eV, which is consistent with the results of $\sigma_1(\omega)$ shown in Figure 3.

Recently, diffusion Monte Carlo calculations showed that the magnetic ground state of the LCO thin film highly depends on the La-La distance, with the HS/LS-FM order becoming the most stable state below a specific compressive strain.[59] In terms of the exchange mechanism, this order mainly results from correlation superexchange, which refers to the magnetic interaction whereby two electrons of $O^{2-}$ mediate the exchange coupling between neighboring HS and LS $Co^{3+}$. More specifically, the double occurrence of AFM superexchange coupling along the HS $Co^{3+}$ $d_{x2-y2}$ – $O^{2-}$ $p_\sigma$ – LS $Co^{3+}$ $d_{x2-y2}$ – $O^{2-}$ $p_\sigma$ – HS $Co^{3+}$ $d_{x2-y2}$ pathway facilitates the FM ordering (Figure S6, Supporting Information). In this mechanism, the net spin moment of LS $Co^{3+}$ $d_{x2-y2}$ (which would be empty in the ideal LS $Co^{3+}$ configuration) may be a descriptor for the superexchange coupling strength because it reflects the spin-polarized charge transfer from neighboring $O^{2-}$ $p_\sigma$ orbitals.

To quantitatively assess the spin-polarized charge transfer, we analyzed the eigen-occupations of the $d$ orbital density matrix of the HS and LS $Co^{3+}$ stabilized in the $c$-units of (100)-, (110)-, and (111)-oriented LCO thin films (Table 2). The eigen-occupations are fictitious auxiliary physical quantities, which often provide important insights into the orbital states of the transition metal ions.[60-62] The HS $Co^{3+}$ was confirmed to have an HS $d^6$ orbital configuration with all majority-spin orbitals and the minority-spin $d_{xy}$ orbital fully occupied, but the other minority-spin orbitals almost empty. The result for LS $Co^{3+}$ shows that the three lower-energy $d$ orbitals, i.e., $d_{xy}$, $d_{yz}$, and $d_{zx}$ orbitals, are almost fully occupied, which is consistent with LS $Co^{3+}$. In contrast, the two higher-energy $d$ orbitals (the $e_g$ orbitals), i.e., $d_{x2-y2}$ and $d_{3z2-r2}$ orbitals, are expected to be empty in the ideal LS $Co^{3+}$ orbital configuration, but as mentioned above, they are partially occupied because of the spin-polarized charge transfer from $O^{2-}$. Interestingly, the net spin moment of LS $Co^{3+}$ $d_{x2-y2}$, estimated by the difference between the eigen-occupations of its majority- and minority-spins, was the largest in the LCO with the (110) orientation (Table 2). This DFT calculation result directly implies that the spin-polarized charge transfer, which determines the superexchange coupling strength, occurs more actively in the (110)-oriented LCO thin films than in the (100)- or (111)- oriented films. In addition, local magnetic moments of HS $Co^{3+}$ within HS/LS-FM order are summarized in Table S3. Upon comparison, the local magnetic moment was the largest in the $c$-unit stabilized the (110)-orientation, followed by that in the (100)- and (111)- orientations, which also matches with experimental results.

**Table 2. Orbital occupations of the HS and LS states for LCO thin films:** A summary for occupations of $t_{2g}$ ($d_{xy}$, $d_{yz}$ and $d_{zx}$) and $e_g$ orbital ($d_{x2-y2}$ and $d_{3z2-r2}$) for HS and LS states.

|  |  |  | HS | | | LS | | |
| --- | --- | --- | --- | --- | --- | --- | --- | --- |
|  |  |  | majority-spin | minority-spin | majority − minority spin | majority-spin | minority-spin | majority − minority spin |
| (100) | $e_g$ | $3z^2-r^2$ | 1.0211 | 0.2333 | 0.7878 | 0.4811 | 0.3258 | 0.1553 |
|  |  | $x^2-y^2$ | 1.0635 | 0.4318 | 0.6317 | 0.4886 | 0.3359 | 0.1527 |
|  | $t_{2g}$ | $xy$ | 0.9726 | 0.9721 | 0.0005 | 0.9737 | 0.9741 | -0.0004 |
|  |  | $yz$ | 0.9943 | 0.2213 | 0.7730 | 0.9801 | 0.9459 | 0.0342 |
|  |  | $zx$ | 0.9932 | 0.2175 | 0.7757 | 0.9797 | 0.9437 | 0.0360 |
| (110) | $e_g$ | $3z^2-r^2$ | 1.0215 | 0.2304 | 0.7911 | 0.4872 | 0.3322 | 0.1550 |
|  |  | $x^2-y^2$ | 1.0601 | 0.4279 | 0.6322 | 0.4806 | 0.3244 | 0.1562 |
|  | $t_{2g}$ | $xy$ | 0.9733 | 0.9730 | 0.0003 | 0.9745 | 0.9747 | -0.0002 |
|  |  | $yz$ | 0.9932 | 0.2154 | 0.7778 | 0.9791 | 0.9466 | 0.0325 |
|  |  | $zx$ | 0.9920 | 0.2167 | 0.7753 | 0.9776 | 0.9435 | 0.0341 |
| (111) | $e_g$ | $3z^2-r^2$ | 1.0250 | 0.2318 | 0.7932 | 0.4957 | 0.3376 | 0.1581 |
|  |  | $x^2-y^2$ | 1.0725 | 0.4331 | 0.6394 | 0.4972 | 0.3502 | 0.1470 |
|  | $t_{2g}$ | $xy$ | 0.9732 | 0.9727 | 0.0005 | 0.9750 | 0.9751 | -0.0001 |

| | | | | | | |
|---|---|---|---|---|---|---|
| yz | 0.9957 | 0.2333 | 0.7624 | 0.9814 | 0.9417 | 0.0397 |
| zx | 0.9957 | 0.2332 | 0.7625 | 0.9814 | 0.9417 | 0.0397 |

## 3. Conclusion

We studied the unexpectedly large modulation of ferromagnetic properties, including a sizable increase in $T_C$, in crystallographic orientation dependent-LCO thin films. By systematically controlling the crystallographic symmetry, we obtained LCO thin films with various structural parameters. The clear FM phases strongly coupled to the crystallographic symmetry demonstrated that the (110)-oriented LCO thin film exhibited highly enhanced FM properties. Optical spectroscopy and DFT calculations revealed that LCO thin films with small $\Delta_{CF}$ and large spin-polarized charge transfer between the Co $3d$ and O $2p$ orbitals facilitate FM ordering in the (110)-oriented LCO thin film. This study implies that the symmetry constraint is an efficient tuning parameter for the exchange coupling strength in ferromagnetic insulator LCO thin films, in which ferromagnetism is highly robust against conventional strain due to its ferroelastic nature.

## 4. Experimental Section

*Epitaxial thin film growth*: High-quality LCO epitaxial thin films were fabricated using PLE on conventionally HF-treated STO substrates with various orientations [(100), (110), and (111)]. An excimer KrF laser ($\lambda = 248$ nm, IPEX864; Lightmachinery) with a fluence of 0.9 J cm$^{-2}$ and a repetition rate of 2 Hz was used. The thin films were synthesized at 500 ºC under an oxygen partial pressure of 100 mTorr. Each set of thin films with distinctive orientations (with the same number of laser shots) was simultaneously fabricated.

*Atomic and crystal structure characterization*: The crystalline structure and lattice parameters of the LCO thin films were characterized using high-resolution XRD (PANalytical X'Pert Pro). HAADF STEM measurements were performed on a Nion UltraSTEM200 microscope operated at 200 kV. The microscope was equipped with a cold field emission gun and a corrector of third- and fifth-order aberrations for sub-angstrom resolution. The collection inner half-angle for the HAADF STEM was 65 mrad. Cross-sectional TEM specimens were prepared via ion milling after conventional mechanical polishing.

*Surface characterization:* Atomic force microscopy measurements were performed using a commercial system (Park Systems, NX10) to examine the surface topography and roughness.

*Magnetization measurements*: A magnetic property measurement system (MPMS3; Quantum Design) was used to characterize the in-plane magnetic properties of the thin films. The temperature-dependent magnetization, $M(T)$, was measured from 300 to 2 K under a magnetic field of 1000 Oe. The magnetic field-dependent magnetization, $M(H)$, was measured at 2 K.

*Chemical composition characterization*: X-ray photoelectron spectroscopy (XPS; AXIS SUPRA, KRATOS Analytical) with Al $K\alpha$ radiation was used to study the chemical state and composition of the LCO thin films. All the X-ray photoelectron spectra were calibrated using the C-C bonding peak (284.5 eV). The Co $2p$ and valence states of the LCO thin films were deconvoluted using Gaussian-Lorentz curves.

*Optical spectroscopy*: The optical conductivity of the LCO thin films was measured using spectroscopic ellipsometry (J. A. Woollam Co., Inc.). A wavelength range from mid-infrared to UV (0.6–6.2 eV) with incident angles of 60° and 65° was used. We employed a three-layer model analysis (surface roughness (50% of material and 50% of voids), LCO, and STO layers) to obtain dielectric functions and optical conductivities of the thin films.

*Theoretical calculations*: *Ab initio* DFT calculations were performed using the Vienna *ab initio* simulation package (VASP) code.[63] The Perdew–Burke–Ernzerhof plus Hubbard correction (PBE + $U$ + $J$) was used for the exchange-correlation functional,[64] in which the double-counting interactions were corrected using the full localized limit (FLL).[65] The values used for the on-site direct Coulomb parameter ($U$) and the anisotropic Coulomb parameter ($J$) were 4.5 and 1.0 eV, respectively.[54] A plane wave basis set at a cutoff energy of 600 eV was used to expand the electronic wave functions, and the valence electrons were described using projector-augmented wave potentials. All atoms were relaxed by the conjugate gradient algorithms until none of the remaining Hellmann–Feynman forces acting on any atoms exceeded 0.02 eV Å$^{-1}$.

**Supporting Information**

Supporting Information is available from the Wiley Online Library or from the author.

**Acknowledgements**

Dongwon Shin and Sangmoon Yoon contributed equally to this work. The authors thank M. F. Chisholm for technical assistance on TEM work. This work was supported by the Basic Science Research Programs through the National Research Foundation of Korea (NRF-2021R1A2C2011340 and NRF-2020K1A3A7A09077715). The microstructural analysis work at ORNL was supported by U.S. DOE, Basic Energy Sciences, Materials Sciences and Engineering Division.

Received: ((will be filled in by the editorial staff))
Revised: ((will be filled in by the editorial staff))
Published online: ((will be filled in by the editorial staff))

Supporting Information

**Tunable Ferromagnetism in LaCoO$_3$ Epitaxial Thin Films**

*Dongwon Shin*[1,#], *Sangmoon Yoon*[2,3,#], *Sehwan Song*[4], *Sungkyun Park*[4], *Ho Nyung Lee*[2], *and Woo Seok Choi*[1,*]

[1]Department of Physics, Sungkyunkwan University, Suwon 16419, Republic of Korea

[2]Materials Science and Technology Division, Oak Ridge National Laboratory, Oak Ridge, Tennessee 37831, United States

[3]Department of Physics, Gachon University, Seongnam, 13120, Republic of Korea

[4]Department of Physics, Pusan National University, Busan 46241, Republic of Korea

[*]Corresponding author: choiws@skku.edu

**Thickness dependent crystal structures of epitaxial LCO thin films**

High-quality LCO thin films were epitaxially grown on STO substrates using PLE as shown in Figure 1a (30 nm), Figure S1a (10 nm) and Figure S1b (20 nm). Figures S1c and 1d show that 10- and 20-nm-thick LCO thin films also exhibit the same orientation-dependent trend in $V_{u.c.}$ and $\varepsilon_{oop}$, as the 30 nm-thick films, as shown in Figures 1c and 1d.

**Absence of oxygen vacancy ordering**

Figure S3a shows HAADF STEM image for the film-interior region that EELS linescan is performed. The EELS spectra of twin-wall and bulk-like domains are compared in Figure S3b and c. First, there is no noticeable difference between two regions at Co-$L_{2,3}$ edge. It is widely known that HS and LS Co$^{3+}$ is hard to be distinguished in Co-$L_{2,3}$ edge because of subtle changes. This results also exclude the presence of oxygen vacancy ordering in pristine twin-wall domains, because the red-shift of Co-$L_3$ edge is not observed in the twin-wall domains. Second, the prepeak (denoted by the black arrow) emerges at the O-$K$ edge spectra of twin-wall domain. It would be the fingerprint that electronic structure, presumably spin states, is different between twin-wall domain and bulk-like domains. However, it is not the direct experimental evidence that ferromagnetism forms in dark-striped regions. Our scenario is supported by theoretical results, and also explains the trend of unconventional magnetic properties. Atomic-scale electron magnetic circular dichroism at low temperature under development may provide direct experimental evidence to end this issue.

**Estimation of spin moment and HS/LS ratio**

To characterize the FM properties of the LCO thin films, we analyzed the $M(T)$ curves using the Curie-Weiss (C-W) law. According to the C-W law, the C-W temperature, spin moment, and ratio of the HS-state were calculated. The inverse of the susceptibility $H/M$ as a function of $T$ above $T_C$ can be fitted by the C-W law $H/M = (T - \theta)/C$, where $C$ is a constant and $\theta$ is the C-W temperature (Figure S4a). All the LCO thin films exhibited a robust FM phase with $\theta$ similar to the $T_C$ values obtained in Figure 2e (Figure S4b). From the $C$ values (slope of the C-W law fitting), we used $\mu = (3k_B C/N_A)^{1/2} \approx 2.84 C^{1/2}$, where $k_B$ is Boltzmann's constant and $N_A$ is Avogadro's number, to determine the spin moment.[1] Furthermore, considering a simple model of $\mu = g\sqrt{S(S+1)}$, where $g \sim 2$ is the gyromagnetic ratio of a free electron and $S$ is the spin value, we determine the ratio of the HS-state, as summarized in Figure S4c.

**Determining Co valence state and the ratio of HS state**

We performed the X-ray photoelectron spectroscopy (XPS) to examine the binding energy of Co, which leads to the estimation of valence state as shown in Figure S5a. Distinct peaks of Co $2p_{3/2}$ and $2p_{1/2}$ are observed without any satellite peaks, indicating the $Co^{3+}$ state within the LCO thin films. The absence of $Co^{2+}$ indicates the absence of oxygen vacancies in the thin films. In addition, the amounts of LS and HS states were estimated. In Figure S5b, the peaks near ~0.235, 0.935, 2.335, and 4.735 eV correspond to the contributions of $Co^{3+}$ HS, LS, O $2p$, and $Co^{3+}$ $3d$, respectively.[2, 3] From the integration of the HS and LS peaks, we obtained the HS (LS) ratio of (100)-, (110)- and (111)-oriented LCO thin films to be 63.07% (36.93%), 65.08% (34.92%), and 61.16% (38.84%), respectively.

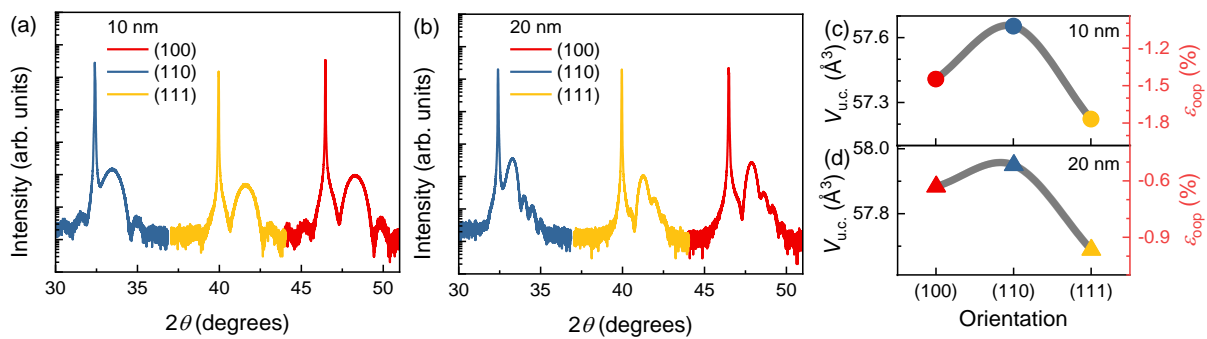

**Figure S1. Crystal structures of the epitaxial LCO thin films (10 and 20 nm) grown on STO substrates.** XRD $\theta$-$2\theta$ scans near the (002), (110), and (111) Bragg planes of a) 10 nm and b) 20 nm LCO thin films (#) on (100), (110), and (111) STO substrates planes (*), respectively. Crystallographic orientation-dependent unit cell volume ($V_{u.c.}$), and out-of-plane strain ($\varepsilon_{oop} = 100 \times (d_{bulk} - d_{film}) / d_{bulk}$) of the c) 10- and d) 20-nm-thick LCO thin films, respectively.

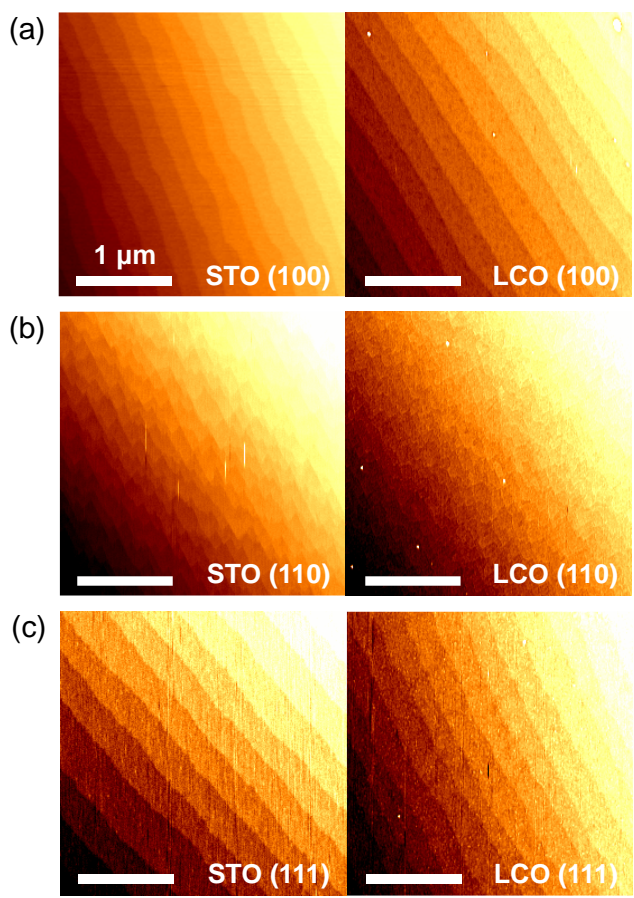

**Figure S2. Surface roughness of STO substrates and LCO thin films.** The rms roughness are 0.118, 0.157, and 0.100 nm for the (100)-, (110)-, and (111)-oriented STO substrates, respectively. In addition, the rms roughness are 0.141, 0.226, and 0.148 nm for (100)-, (110)-, and (111)-oriented LCO thin films, respectively.

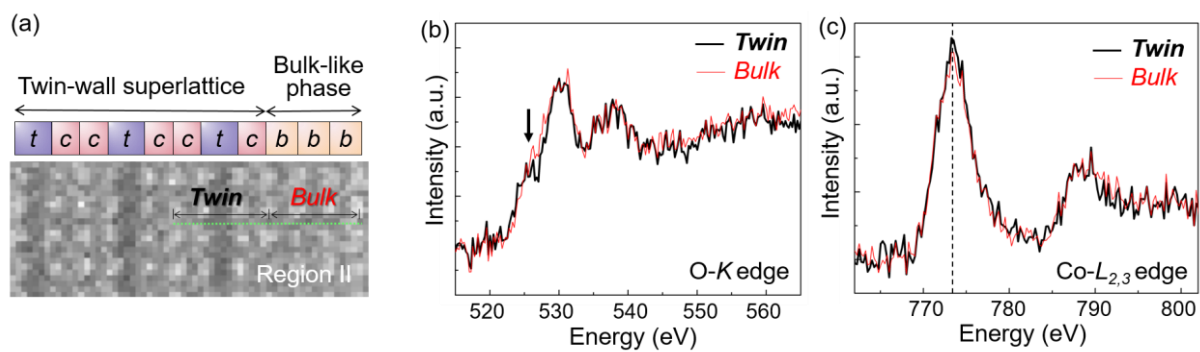

**Figure S3. STEM-EELS (electron energy loss spectroscopy) results.** a) HAADF STEM image for the film-interior region that EELS linescan is performed. EELS b) O $K$-edge and c) Co $L$-edge spectra are respectively shown for the twin-wall and bulk-like domains.

| c-unit | (100) | (110) | (111) |
|---|---|---|---|
| 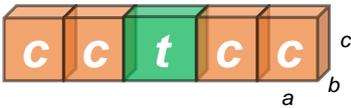 | 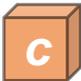 | 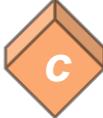 | 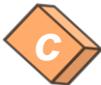 |
| a [Å] | 3.86 | 3.85 | 3.85 |
| b [Å] | 3.905 | 3.905 | 3.85 |
| c [Å] | 3.68 | 3.7 | 3.62 |
| V [Å³] | 55.47 | 55.63 | 53.66 |

**Table S1. Structural properties of the *c*-units with different crystallographic symmetry in the FM state.** Lattice parameters of *a*, *b*, *c*, and *V* of the *c*-units, shown for LCO thin films with different crystallographic orientations. The schematics show distorted *c*-units by the crystallographic symmetry modulation.

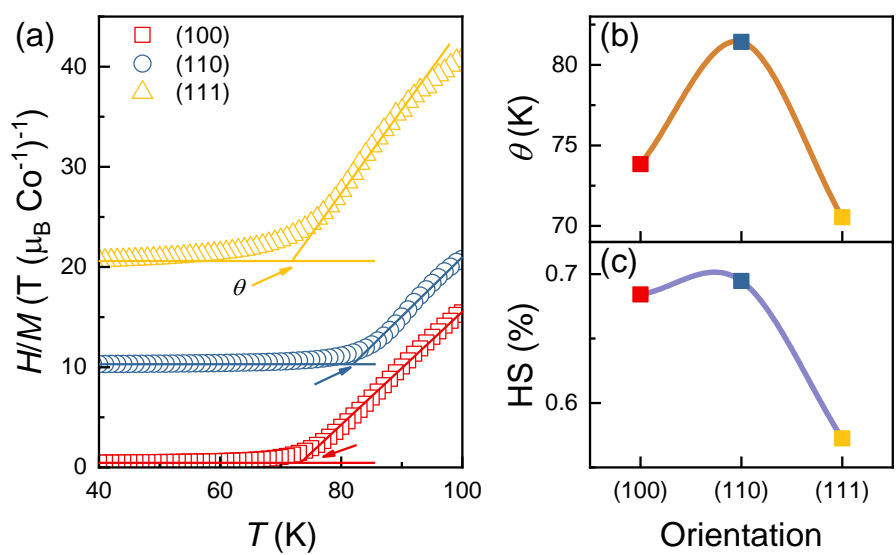

**Figure S4. FM properties of LCO thin films from the Curie-Weiss law fitting.** a) The inverse of the susceptibility $H/M$ as a function of $T$ above $T_C$. The solid lines indicate Curie-Weiss law fitting $H/M = (T - \theta) / C$, where $C$ is constant, $\theta$ is the C-W temperature. The temperature-dependent magnetic susceptibility provides spin moment and ratio of the HS state estimated by the C-W law. The summary of the orientation-dependent FM properties; b) $\theta$ and c) HS state ratio.

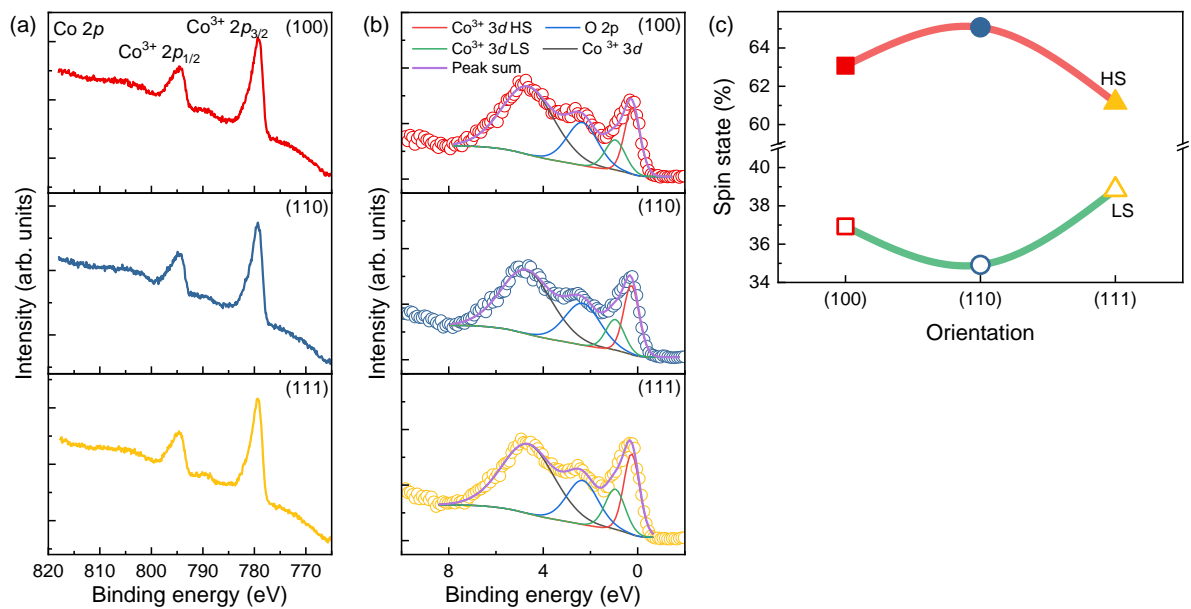

**Figure S5. Characterization of Co valence state and spin states.** a) XPS spectra of Co 2$p$ levels. b) XPS spectra near the valence-band region. c) Summary of the LS and HS ratio in the LCO thin films with distinct crystallographic orientations.

| | | AFM (HS) | FM (HS/LS) | FM (HS) | NM (LS) |
|---|---|---|---|---|---|
| Energy [meV/f.u.] | (100) | 0 | -55 | 274 | 297 |
| | (110) | 0 | -50 | -25 | 277 |
| | (111) | 0 | -129 | 227 | 185 |

**Table S2. Total energy calculations of the $c$-units with different symmetry in four different magnetic configurations.** The magnetic configurations are schematically demonstrated in the first raw of the table, where HS (orange sphere) and LS (blue sphere) of Co atoms and spins (black arrow) are shown. The energy of each magnetic configuration is represented relative to that of the HS AFM configuration, which is set to zero energy.

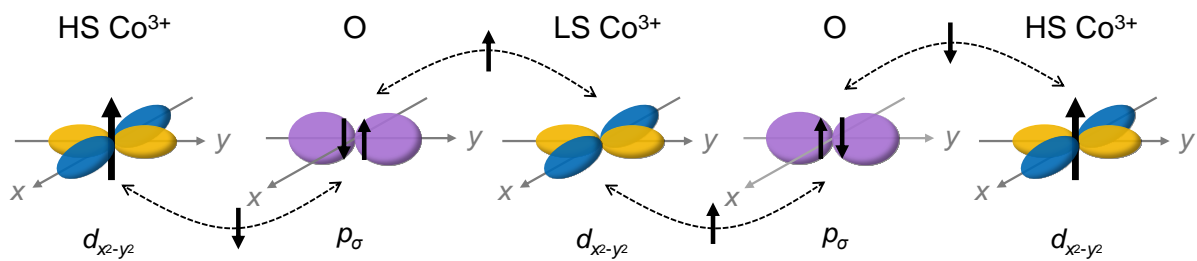

**Figure S6. Proposed FM superexchange mechanism for the epitaxial LCO thin film.** Schematic representation of FM HS $Co^{3+}$ - $Co^{3+}$ superexchange interaction promoted via empty LS $Co^{3+}$ and filled O 2p orbitals. HS $Co^{3+}$ $d_{x2-y2}$ (blue and yellow lobes), $O^{2-}$ $p_\sigma$ (purple lobes) and spin (black arrow) are shown.

|  | (100) | (110) | (111) |
|---|---|---|---|
| $M^{HS\ Co^{3+}}$ ($\mu_B$) | 2.986 | 2.993 | 2.976 |

**Table S3. The local magnetic moment of $Co^{3+}$ within the *c*-units.**